\documentclass[prl,twocolumn]{revtex4}

\usepackage{graphicx}% Include figure files

\usepackage{amssymb}

\def \nn{\nonumber}

\begin{document}

\title{Nernst effect and diamagnetism in phase fluctuating superconductors}
\author{Daniel Podolsky$^{1,2}$, Srinivas Raghu$^{3,4}$, and Ashvin Vishwanath$^{1,2}$}
\affiliation{$^1$Department of Physics, University of California, Berkeley, CA 94720\\
$^2$Materials Sciences Division, Lawrence Berkeley National Laboratory, Berkeley, CA 94720\\
$^3$Department of Physics, Princeton University, Princeton, NJ
08544\\
$^4$Physics Department, Stanford University, Stanford, CA 94305}
\date{Printed \today}

%\begin{affiliations}
% \item Department of Physics, University of California, Berkeley, CA 94720
% \item Materials Sciences Division, Lawrence Berkeley National Laboratory, Berkeley,
%CA 94720
% \item Department of Physics, Princeton University, Princeton, NJ
% 08544

% \item Physics Department, Stanford University, Stanford, CA 94305

%\end{affiliations}

\begin{abstract}

When a superconductor is warmed above its critical temperature
$T_c$, long range order is destroyed by fluctuations in the order
parameter. These fluctuations can be probed by measurements of
conductivity\cite{Orenstein}, diamagnetism\cite{ong1} and of the
Nernst effect\cite{ong2,behnia1,behnia2}. Here, we study a regime
where superconductivity is destroyed by phase fluctuations arising
from a dilute liquid of mobile vortices. We find that the Nernst
effect and diamagnetic response differ significantly from Gaussian
fluctuations -- in particular, a much sharper decay with
temperature is obtained. We predict a rapid onset of Nernst signal
at a temperature T$_{\rm onset}$ that tracks $T_c$, rather than
the pairing temperature. We also predict a close quantitative
connection with diamagnetism -- the ratio of magnetization to
transverse thermoelectric conductivity $\alpha_{xy}$ reaches a
universal value at high temperatures. We interpret Nernst effect
measurements on the underdoped cuprates in terms of a dilute
vortex liquid over a wide temperature range above $T_c$.

\end{abstract}

\maketitle

In recent years, the Nernst effect has emerged as an important probe
of strongly-correlated electron systems.
%The experiment consists of
%applying a temperature gradient and a magnetic field in
%perpendicular directions (say, along $\hat{x}$ and $\hat{z}$), and
%measuring the electric field in the third direction, with open
%circuit boundary conditions.
The Nernst signal is the electric field ($E_y$) response to a
transverse temperature gradient $\nabla_x T$,
\begin{equation}
e_N \equiv  \left.\frac{E_y}{-\nabla_x T}\right|_{J=0}
\end{equation}
A large Nernst signal has been detected in the normal state of
quasi-2d cuprate\cite{ong2} and heavy fermion\cite{behnia1}
samples, and in thin films\cite{behnia2}, well above $T_c$. This
is in contrast with typical (non-ambipolar) metals, where the
Nernst effect is usually weak: it was shown\cite{sondheimer} that
Fermi liquid quasiparticles with energy-independent scattering
rates do not contribute to the Nernst signal. This, together with
the proximity of the large-Nernst region in some of these
materials to superconducting phases, points towards fluctuating
superconductivity as one natural source for the Nernst signal.

Theoretical studies of the Nernst effect in cuprate
superconductors include the analysis of Gaussian fluctuations
above the mean-field transition temperature\cite{Ussishkin1}, and
a Ginsburg-Landau model with interactions between fluctuations of
the order parameter\cite{Mukerjee1}. In this Letter, we consider
the Nernst effect due to thermal fluctuations in a phase-only
model. Throughout, we assume that the superconducting order
parameter $\psi(x)=\Delta_0 e^{i\theta(x)}$ has a frozen amplitude
$\Delta_0(x)=const$. Such a situation can arise in granular thin
films and in Josephson junction arrays, where the order parameter
on individual grains is well-established, whereas $T_c$ is given
by the weak Josephson coupling between grains. This picture may
also be appropriate for the underdoped cuprates, where the pairing
gap is thought to be much larger than the transition temperature,
$k_BT_c\ll\Delta$\cite{EmeryKivelson,FisherFisherHuse}. Here,
superconductivity is destroyed by loss of phase coherence via
thermally generated vortex anti-vortex pairs. Phase fluctuations
from vortex diffusion have been proposed as the dominant
contribution to the Nernst signal\cite{ong2,anderson}. This vortex
picture is most useful in the dilute limit, when the spacing
between vortices (both field induced and thermally generated) is
much larger than their core radius  (the zero temperature
coherence length $\xi_0$). Then, vortices have a well defined
identity and the amplitude is suppressed only in the small area of
the sample occupied by vortex cores.

Our starting point is the Lawrence-Doniach model of a layered
superconductor,
\begin{eqnarray}
F_{LD}&=&-{\mathcal J}\sum_n\sum_{\langle ij\rangle}\left \{
\psi^*_{j,n}e^{{\rm i}A_{ij}}\psi_{i,n}+ {\rm h.c.}\right
\}\nn\\&\,&-{\mathcal J_\perp}\sum_{\langle nm\rangle}\sum_i \left
\{ \psi^*_{i,n}\psi_{i,m}+ {\rm h.c.}\right \}\nn\\&\,&
+U\sum_n\sum_{i}(|\psi_{i,n}|^2+r/2U)^2 \label{eq:Fld}
\end{eqnarray}
where $i,j$ label lattice points within a layer, and $n,m$ label
the layers.  The lattice vector potential due to an external
magnetic field, $A_{ij}=\frac{2e}{\hbar}\int_{{\bf r}_i}^{{\bf
r}_j} d{\bf r}\cdot {\bf A}$, is static and unscreened,
corresponding to an extreme type-II superconductor. We consider
the limit deep in the ordered phase within mean field theory $-r
\gg k_BT$, where phase fluctuations dominate $\psi_{i,n} =
\Delta_0 e^{i\theta_{i,n}}$. This reduces the model above to an
${\rm XY}$ model. The inter-layer coupling ${\mathcal J_\perp}$
stabilizes true long-range superconductivity. However, we have
verified that realistic values of ${\cal J}_\perp$ increase $T_c$
relative to the 2D Kosterlitz-Thouless transition $T_{KT}$ only by
a small amount, and that the normal state properties of interest
are not significantly affected by ${\mathcal J_\perp}$, except
very close to $T_c$. Hence, in what follows we set ${\mathcal
J_\perp}=0$ and consider the 2D ${\rm XY}$ model with Josephson
coupling $J=\Delta_0^2 {\cal J}$.

To study transport, we supplement this statistical mechanics model
with model-A Langevin dynamics, corresponding to interaction with
a heat bath that does not preserve any conservation
laws\cite{HohenbergHalperin}:
\begin{eqnarray}
F_{\rm XY} &=& - J\sum_{\langle ij \rangle} \cos(\theta_i -\theta_j -A_{ij})  \nn \\
 \tau\partial_t\theta_{i}&=&-\frac{\partial F_{\rm
XY}}{\partial \theta_{i}}+\eta_{i}(t)\label{eq:Lang}.
\end{eqnarray}
Here, $\tau$ provides a characteristic time scale for the
dynamics. The stochastic noise $\eta_i(t)$ is Gaussian correlated,
with a variance chosen to satisfy the fluctuation-dissipation
theorem,
\begin{eqnarray}
\langle \eta_{i}(t)\eta_{j}(t')\rangle &=& 2k_BT\tau\,
\delta_{ij}\delta(t-t').\label{eq:noiseCorr}
\end{eqnarray}
The model (\ref{eq:Lang}), and (\ref{eq:noiseCorr}) has only three
free parameters: $J$, $\tau$, and the lattice constant, $a$ (or
equivalently a field scale, $H_{0}=\Phi_0/(2\pi a^2$), defined via
the superconducting flux quantum $\Phi_0$).  Of these, $J$ is an
overall energy scale, set by fixing $T_{KT}$. The length scale $a$
depends on the physical system in question. For Josephson junction
arrays or granular superconductors, this is the spacing between
grains. For uniform superconductors, $a$ can be determined by
comparing the correlation length in the XY model away from the
transition, eg. at $T=2T_c$, with the typical separation between
thermally induced vortices at that temperature. Thus, $a$ is given
by a combination of vortex fugacity and core radius
$\xi_0$\cite{HonerkampLee}.  Within $\xi_0$, the superconducting
amplitude is significantly suppressed.  Hence, the dilute limit,
where the separation between vortices exceeds $\xi_0$, determines
the temperature window over which a phase-only description is
appropriate. Finally, the time scale $\tau$ does not affect
thermodynamic quantities, such as magnetization, nor does it enter
the transverse thermoelectric conductivity, $\alpha_{xy}$, which
is closely related to the Nernst effect,
\begin{eqnarray}
e_N =\frac{\alpha_{xy}}{\sigma_{xx}},
\end{eqnarray}
where $\sigma_{xx}$ is the electrical conductivity, and we have
assumed particle-hole symmetry.  Hence, $\alpha_{xy}$ and
magnetization predicted by this model are only functions of
$T/T_{KT}$ and $H/H_0$.

By an Onsager relation, the transverse thermoelectric conductivity
$\alpha_{xy}$ can be obtained either from the electric current
response to a temperature gradient, $j_{{\rm
tr},x}=-\alpha_{xy}\nabla_y T$, or from the heat current response
to an electric field, $j^Q_{{\rm tr},x}=T\alpha_{xy} E_y$.  A
third method to compute $\alpha_{xy}$ is through a Kubo formula
involving the unequal time correlator $\langle
j_x(t)j_y^Q(0)\rangle$. As a check of our numerics, and of the use
of proper magnetization subtractions\cite{CooperHalperinRuzin}, we
confirmed that all three methods agree for a representative set of
temperatures and magnetic fields.

\begin{figure}
\includegraphics[width=9cm]{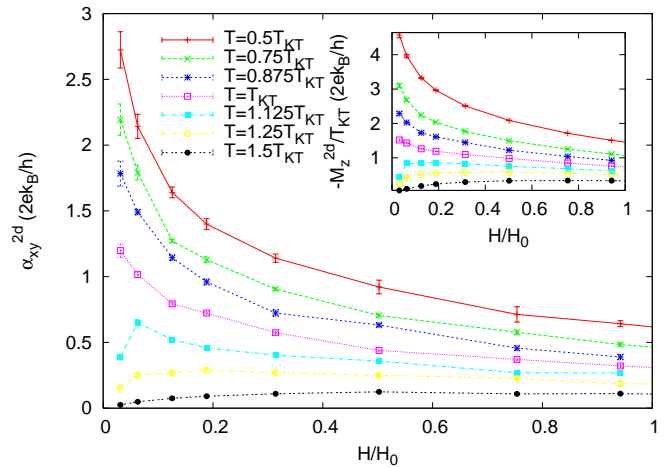}
\caption{Transverse thermoelectric conductivity for a single
plane, in units of the quantum of thermoelectric conductance
$2ek_B/h$.  Simulations on a cylindrical geometry, with system
size ranging from $60\times 60$ to $200\times 200$. {\it Inset:}
Diamagnetic response. Temperatures as in the main figure.
\label{fig:AvsH}}
\end{figure}

\begin{figure}
\includegraphics[width=9cm]{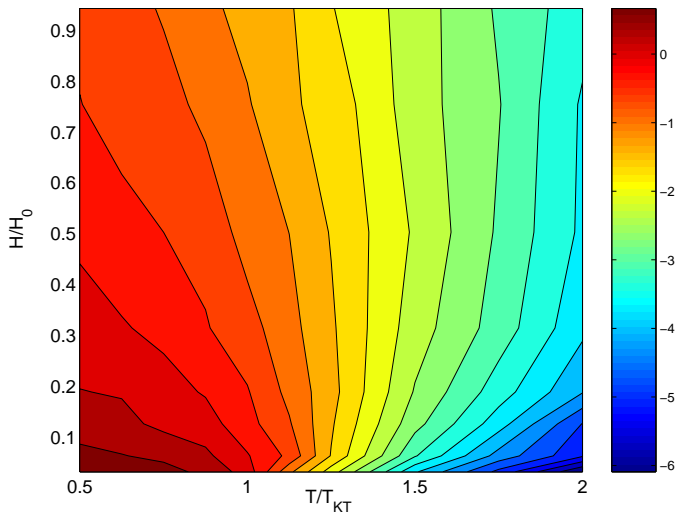}
\caption{Contour plot of
$\log\left[\alpha_{xy}^{2d}/(2ek_B/h)\right]$. For small $T$ and
$H$, an inter-layer Josephson coupling $J_\perp$, absent in these
simulations, stabilizes 3d superconductivity.  Note that, unlike
the Nernst signal $e_N=\alpha_{xy}/\sigma_{xx}$ in
Ref.~\cite{ong2}, there is no ridge field in our simulations of
$\alpha_{xy}$.  The difference is likely due to the diverging
electrical conductivity $\sigma_{xx}$ as $T_c$ is approached.
\label{fig:Acont}}
\end{figure}

{\em Intermediate Temperature Regime:} The results for
$\alpha^{2d}_{xy}$ on a single layer are shown in
Figs.~\ref{fig:AvsH} and~\ref{fig:Acont}. As noted above,
$\alpha_{xy}$ is independent of the parameter $\tau$. For a single
layer, its value only depends on the ratios $T/T_{KT}$ and
$H/H_0$, and is expressed naturally in units of the 2D ``quantum
of thermoelectric conductance", $2 e k_B/h$.  To compare with thin
film and layered systems, one must divide by the film
thickness/layer separation $d$, $\alpha_{xy}=\alpha^{2d}_{xy}/d$.
The inset of Fig.~\ref{fig:AvsH} shows the diamagnetic response.
Below $T_c$, the magnetization diverges logarithmically in $H$.
The magnetization edge currents correspond to a ``depletion layer"
near the edge of the sample, where vortex density is smaller than
the density in the bulk, $H/\Phi_0$.   In the vortex picture,
magnetization is analogous to the work function in a metal, and we
find\cite{RaghuPodolskyVishwanath}, to leading order in $H$,
\begin{eqnarray}
M_z^{2d}=-\frac{2e}{h}\left(\frac{\pi\rho_s}{2}-k_BT\right)\log
\frac{H_0}{H}\label{eq:mag}
\end{eqnarray}
This is similar to a previous result\cite{OganesyanHuseSondhi},
and is in good agreement with our simulations.

\begin{figure}
\includegraphics[width=9cm]{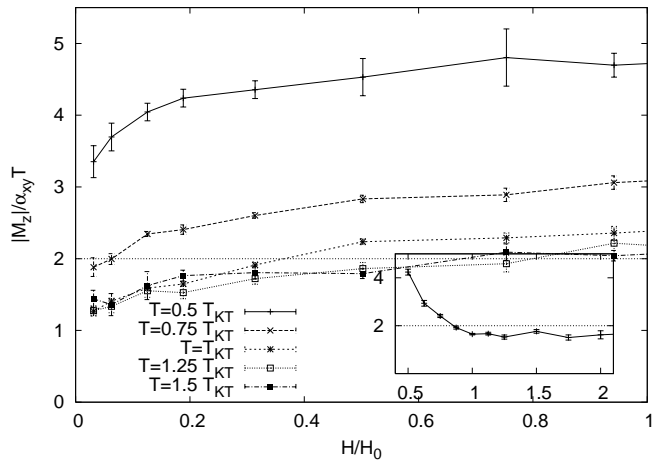}
\caption{The dimensionless ratio $|M_z|/T\alpha_{xy}$, as a
function of magnetic field for various temperatures.  In the
high-temperature limit, this dimensionless ratio is expected to
saturate at a value of 2 for all magnetic fields. {\it Inset:}
$|M_z|/T\alpha_{xy}$ for $H=0.31 H_{0}$ vs.
$T/T_{KT}$\label{fig:MoA}}
\end{figure}

{\em High Temperature Expansion:}  For $T\ll J$, the phase-only
model allows for an analytically tractable regime that is entirely
different from the Gaussian regime considered
previously\cite{Ussishkin1}. The high temperature expansion,
carried out in powers of $J/k_BT$, is conveniently performed using
the Martin-Siggia-Rose
formalism\cite{MartinSiggiaRose,RaghuPodolskyVishwanath}. Since
both $M_z$ and $\alpha_{xy}$ require a magnetic field, the
expansion of these quantities involves graphs enclosing finite a
magnetic flux. The leading term thus depends on the smallest
closed graph -- on a square lattice this involves 4 links, and is
hence proportional to $(J/T)^4$, whereas on a triangular lattice
it goes as $(J/T)^3$:
\begin{eqnarray}
\alpha^{2d}_{xy}&=&\lambda\frac{2ek_B}{h}\left(\frac{J}{T}\right)^\mu\sin\frac{H}{H_0}\label{eq:largeT}\\
\frac{M_z^{2d}}{T}&=&-2\lambda\frac{2ek_B}{h}\left(\frac{J}{T}\right)^\mu\sin\frac{H}{H_0}\nonumber\\
\frac{|M_z^{2d}|}{T\alpha_{xy}^{2d}}&=&2.\label{eq:MoA}
\end{eqnarray}
Here, $\mu=4$ and $\lambda=\pi/8$ ($\mu=3$ and $\lambda=\pi/4$)
for a square (triangular) lattice.  Despite the lattice-dependent
behavior of $\alpha_{xy}$ and $M_{z}/T$, their ratio
(\ref{eq:MoA}) is equal to -2, independent of the lattice.  The
same value of -2 is obtained in the Gaussian regime of the
dynamical Ginzburg-Landau equation\cite{Ussishkin2,Ussishkin1},
and seems to be a robust feature of fluctuating superconductivity
at high temperatures. The ratio $|M_z|/T\alpha_{xy}$, shown in
Fig.~\ref{fig:MoA}, is only weakly field-dependent, and tends to 2
at high-temperatures.  This points to a close quantitative
connection between the Nernst effect and diamagnetism \cite{ong1}.

\begin{figure}
\includegraphics[width=9cm]{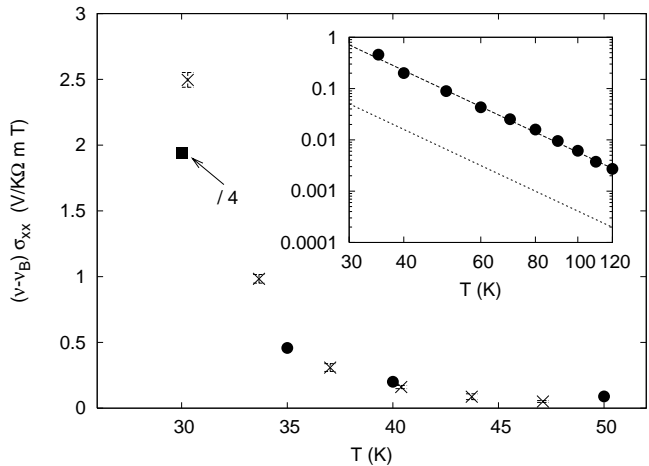}
\caption{Comparison to underdoped La$_{2-x}$Sr$_x$CuO$_4$
($x=0.12$, $T_c=28$ K). The experimental data\cite{ong2}, shifted
by a constant quasiparticle background contribution
$\nu_B\sigma_{xx}=-0.011 V/K\Omega Tm$, is indicated by $\bullet$.
Simulation results are shown by X with error bars. The value of
the experimental point at $T=30 K$ ($\blacksquare$) has been
divided by 4 to fit in the figure. This point is very close to
$T_c$, and hence $\nu\sigma_{xx}$ is likely dominated by 3d
fluctuations, not considered in our simulations. \label{fig:yayu}}
\end{figure}

{\em Comparison to Nernst Measurements in the Cuprates:}
Experimental measurements of both Nernst voltage and conductivity
are required to obtain $\alpha_{xy}$. Such experimental data is
available on underdoped La$_{2-x}$Sr$_x$CuO$_4$ ($x=0.12$ and
$T_c=28$ K) in weak fields. This is shown in Fig.~\ref{fig:yayu},
which displays the Nernst coefficient times conductivity,
$\nu\sigma_{xx}=\left.\frac{d\alpha_{xy}}{dH}\right|_{H=0}$. To
compare, we choose in our simulation, $J=J_l\equiv30.2$ K,
corresponding to $T_c=1.04 T_{KT}$, and $H_0=50$ T (for this sample,
$H_{c2}\approx$100 T, hence $H_{c2}>H_0$, consistent with a
relatively dilute vortex liquid). With these values, the simulation
gives good agreement with absolute experimental values in the regime
$T_c<T<2T_c$ K, except for the very lowest temperature point $T=30$
K. This is very close to $T_c$, so that 3d superconducting
fluctuations, ignored here, are likely dominant.

{\em Comparison with High Temperature Data:} The inset of
Fig.~\ref{fig:yayu} shows the measured $\nu\sigma_{xx}$ on a
log-log plot extending to $T=120 K\approx 4T_c$.  The data
displays a rapid decay over a large temperature range, in general
agreement with our expectations. In particular, a $T^{-4}$ decay
is observed, which is the high temperature result
(\ref{eq:largeT}) on a square lattice.  However, since in the high
temperature regime the precise power depends strongly on lattice
geometry (in contrast to the intermediate temperature regime)
justification for using a nearest neighbor square lattice model
(as opposed to, say, a triangular lattice model) is required.
While the underlying $d$-wave symmetry of the cuprates, coupled
with the fact that $a$ is a microscopic length, about 6 times the
lattice spacing, may be invoked, whether these are sufficient to
justify the square lattice model is unclear at present and
requires further work.

The characteristic scale of $\alpha_{xy}$ at these temperatures is
enhanced from what one would expect for a superconductor with
$T_c=28K$. For example, the best fit to the square lattice high
temperature expansion (solid line) requires $J=J_h=52 K$, larger
than the effective coupling $J_l=30.2$ K that yiels the correct
$T_c$. This may be naturally attributed to thermal d-wave
quasiparticles, omitted in this analysis, which suppress the long
distance superfluid density\cite{LeeWen} but not the superfluid
density at shorter scales\cite{RaghuPodolskyVishwanath}, which
controls the high temperature behavior. The ratio $J_h/J_l=1.6$ is
consistent with measurements of the temperature dependent
superfluid density in other cuprates \cite{lemberger}. A
prediction from this scenario is that magnetization should
continue to track $\alpha_{xy}T$.

{\em Onset Temperature:} Ong and collaborators\cite{ong2} define a
temperature $T_{\rm onset}$ where the fluctuating contribution to
the Nernst effect can no longer be experimentally distinguished
from the quasiparticle background. Here, since we have a natural
scale for $\alpha_{xy}$, we define $T_{\rm onset}$ as the
temperature where $\alpha_{xy}^{2d}$ has decayed to a small
fraction $\delta$ of the quantum of thermoelectric conductivity,
\begin{eqnarray}
\alpha_{xy}^{2d}(T_{onset},H)=\frac{2ek_B}{h}\,\delta
\end{eqnarray}
For our model,
$\alpha_{xy}^{2d}=\frac{2ek_B}{h}F\left(T/T_{KT},H/H_0\right)$,
hence $T_{\rm onset}$ is proportional to $T_c$.  The essential
point is that, because $\alpha_{xy}^{2d}$ depends strongly on
temperature, when inverted, $T_{onset}$ is only a weak function of
$\delta$ and $H$.  For instance, the choice $H=H_0/4$ and
$\delta=0.01$ on the square lattice yields
\begin{eqnarray}
T_{\rm onset}\approx 3 T_c
\end{eqnarray}
This is consistent with the observation that the
experimentally-defined $T_{\rm onset}$ roughly tracks $T_c$ as
doping is varied\cite{ong2}.  Due to the strong temperature
dependence of $\alpha_{xy}$, the onset of Nernst effect is very
sharp, in contrast to Gaussian fluctuations\cite{Ussishkin1}, where
$\alpha_{xy}$ only decays as $1/(T_c-T)$ at high temperatures.

Experimentally, measurements of the electrical conductivity
$\sigma_{xx}$ do not have discernable contributions due to
fluctuating superconductivity at temperatures of order $T_{\rm
onset}$\cite{Orenstein}. Since $\sigma_{xx}$ is proportional to the
parameter $\tau$ appearing in Eq.~(\ref{eq:Lang}), this places a
constraint on the maximum value of $\tau$.  The high temperature
expansion yields
\begin{eqnarray}
\sigma^{2d}_{xx,\rm fluct}=\frac{4e^2}{h}\frac{\pi J_{\rm
eff}^2}{4T^2}\frac{\tau}{\hbar}.\nn
\end{eqnarray}
As a benchmark, we note that BCS theory predicts the value
$\tau_{\rm BCS}\approx 0.7 \hbar$. This yields a fluctuating
conductivity at $T=2T_c$ that is only about $10\%$ of the
quasiparticle conductivity in this material.

In conclusion, we have studied the transverse thermoelectric
conductivity $\alpha_{xy}$ and the diamagnetic response $M^z$ in
the classical XY model with model-A dynamics. We have obtained
numerical results at low temperatures, and analytic results at
high temperatures, that are functions only of two variables,
$T/T_c$ and $H/H_0$, where $H_0<H_{c2}$ is a characteristic field
scale set by vortex parameters.  In our model, we predict that
$\alpha_{xy}$ and $M^z$ for different systems (e.g. different
dopings) should collapse into a single curve when expressed in
terms of the system-dependent $T_c$ and $H_0$. We show that
$M_z/T$ and $\alpha_{xy}$ track each other and, in particular, we
predict that their ratio tends to $-2$ at high temperatures.
Measurements of $\alpha_{xy}$ on the underdoped cuprate
La$_{1.88}$Sr$_{0.12}$CuO$_4$ display a sharp temperature decay,
in agreement with our model.

We would like to thank P.W. Anderson, D. Huse, S. Kivelson, J.E.
Moore, S. Mukerjee, N.P. Ong, and Y. Wang for many insightful
discussions.

\end{document}